\documentclass{article}
\usepackage{amsmath,amssymb,url,amsthm,mathpazo,color}
\usepackage{graphicx}
\usepackage{type1cm}
\usepackage{subcaption}
\usepackage{geometry}
\newtheorem{thm}{Theorem}[section]

\newtheorem{lem}[thm]{Lemma}
\newtheorem{prop}[thm]{Proposition}
\newtheorem{defn}[thm]{Definition}
\newtheorem{rem}[thm]{Remark}
\newtheorem{ass}[thm]{Assumption}

\allowdisplaybreaks
\def\fin   {\hfill{$\Box$}\vspace{5mm}}
\def\l     {\left}
\def\r     {\right}
\def\<     {\langle}
\def\>     {\rangle}

\def\calB  {{\cal B}}
\def\calE  {{\cal E}}
\def\calF  {{\cal F}}
\def\bbC   {{\mathbb C}}
\def\bbE   {{\mathbb E}}
\def\bbF   {{\mathbb F}}

\def\bbP   {{\mathbb P}}

\def\bbR   {{\mathbb R}}
\def\ve    {\varepsilon}
\def\vp    {\varphi}
\def\vt    {\vartheta}
\def\tP    {\bbP^\ast}
\def\tN    {\widetilde{N}}
\def\theequation{\thesection.\arabic{equation}}
\begin{document}
\title{Numerical analysis on quadratic hedging strategies for normal inverse Gaussian models\footnote{This work was supported by JSPS KAKENHI Grant Number 15K04936 and 17K13764.}}
\author{Takuji Arai\footnote{Department of Economics, Keio University, Email: arai@econ.keio.ac.jp}, 
        Yuto Imai\footnote{Mitsubishi UFJ Trust Investment Technology Institute Co., Ltd.(MTEC), and Research Institute for Science and Engineering, Waseda University}
        and Ryo Nakashima\footnote{Department of Economics, Keio University}}
\maketitle

\begin{abstract}
The authors aim to develop numerical schemes of the two representative quadratic hedging strategies: locally risk minimizing and mean-variance hedging strategies,
for models whose asset price process is given by the exponential of a normal inverse Gaussian process, using the results of Arai et al. \cite{AIS}, and Arai and Imai \cite{AI}.
Here normal inverse Gaussian process is a framework of L\'evy processes frequently appeared in financial literature.
In addition, some numerical results are also introduced.

\vspace{3mm}

\end{abstract}

%
%
\setcounter{equation}{0}
\section{Introduction}
Locally risk minimizing (LRM) and mean-variance hedging (MVH) strategies are well-known quadratic hedging strategies for contingent claims in incomplete markets.
In fact, their theoretical aspects have been studied very well for about three decades.
On the other hand, numerical methods to compute them have yet to be thoroughly developed.
As limited literature, Arai et al. \cite{AIS} developed a numerical scheme of LRM strategies for call options for two exponential L\'evy models:
Merton jump-diffusion models and variance Gamma (VG) models.
Here VG models mean models in which the asset price process is given as the exponential of a VG process.
In \cite{AIS}, they made use of a representation for LRM strategies provided by Arai and Suzuki \cite{AS}, and the so-called Carr-Madan method suggested by \cite{CM}:
a computational method for option prices using the fast Fourier transforms (FFT).
Note that \cite{AS} obtained their representation for LRM strategies by means of Malliavin calculus for L\'evy processes.
As for MVH strategies, Arai and Imai \cite{AI} obtained a new closed-form representation for exponential additive models, and suggested a numerical scheme for VG models.

Our aim in this paper is to extend the results of \cite{AIS} and \cite{AI} to normal inverse Gaussian (NIG) models.
Note that an NIG process is a pure jump L\'evy process described as a time-changed Brownian motion as well as a VG process is.
Here a process $X=\{X_t\}_{t\geq0}$ is called a time-changed Brownian motion, if $X$ is described as
\[
X_t=\mu Y_t+\sigma B_{Y_t}
\]
for any $t\geq0$, where $\mu\in\bbR$, $\sigma>0$, $B=\{B_t\}_{t\geq0}$ is a one-dimensional standard Brownian motion, and $Y=\{Y_t\}_{t\geq0}$ is a subordinator,
that is, a nondecreasing L\'evy process.
A time-changed Brownian motion $X$ is called an NIG process, if the corresponding subordinator $Y$ is an inverse Gaussian (IG) process.
On the other hand, a VG process is described as a time-changed Brownian motion with Gamma subordinator.
NIG process, which has been introduced by Barndorff-Nielsen \cite{BN}, is frequently appeared in financial literature,
e.g. \cite{BN0}, \cite{BN1}, \cite{BB}, \cite{R1}, \cite{R2} and so forth.

Next, we introduce quadratic hedging strategies.
Consider a financial market composed of one risk-free asset and one risky asset with finite maturity $T>0$.
For simplicity, we assume that market's interest rate is zero, that is, the price of the risk-free asset is 1 at all times.
Let $S=\{S_t\}_{t\in[0,T]}$ be the risky asset price process.
Here we prepare some terminologies.

\begin{defn}
\begin{enumerate}
\item A strategy is defined as a pair $\vp=(\xi, \eta)$, where $\xi=\{\xi_t\}_{t\in[0,T]}$ is a predictable process, and $\eta=\{\eta_t\}_{t\in[0,T]}$ is an adapted process.
      Note that $\xi_t$ (resp. $\eta_t$) represents the amount of units of the risky asset (resp. the risk-free asset) an investor holds at time $t$.
      The wealth of the strategy $\vp=(\xi, \eta)$ at time $t\in[0,T]$ is given as $V_t(\vp):=\xi_tS_t+\eta_t$.
      In particular, $V_0(\vp)$ gives the initial cost of $\vp$.
\item A strategy $\vp$ is said to be self-financing, if it satisfies $V_t(\vp)=V_0(\vp)+G_t(\xi)$ for any $t\in[0,T]$,
      where $G(\xi)=\{G_t(\xi)\}_{t\in[0,T]}$ denotes the gain process induced by $\xi$, that is, $G_t(\xi):=\int_0^t\xi_udS_u$ for $t\in[0,T]$.
      If a strategy $\vp$ is self-financing, then $\eta$ is automatically determined by $\xi$ and the initial cost $V_0(\vp)$.
      Thus, a self-financing strategy $\vp$ can be described by a pair $(\xi,V_0(\vp))$.
\item For a strategy $\vp$, a process $C(\vp)=\{C_t(\vp)\}_{t\in[0,T]}$ defined by $C_t(\vp):=V_t(\vp)-G_t(\xi)$ for $t\in[0,T]$
      is called the cost process of $\vp$.
      When $\vp$ is self-financing, its cost process $C(\vp)$ is a constant.
\item Let $F$ be a square integrable random variable, which represents the payoff of a contingent claim at the maturity $T$.
      A strategy $\vp$ is said to replicate claim $F$, if it satisfies $V_T(\vp)=F$.
\end{enumerate}
\end{defn}

\noindent
Roughly speaking, a strategy $\vp^F=(\xi^F,\eta^F)$, which is not necessarily self-financing, is called the LRM strategy for claim $F$,
if it is the replicating strategy minimizing a risk caused by $C(\vp^F)$ in the $L^2$-sense among all replicating strategies.
Note that it is sufficient to get a representation of $\xi^F$ in order to obtain the LRM strategy $\vp^F$, since $\eta^F$ is automatically determined by $\xi^F$.
On the other hand, the MVH strategy for claim $F$ is defined as the self-financing strategy minimizing the corresponding $L^2$-hedging error,
that is, the solution $(\vt^F, c^F)$ to the minimization problem
\[
\min_{c,\vt}\bbE\l[\l(F-c-G_T(\vt)\r)^2\r].
\]
Remark that $c^F$ gives the initial cost, which is regarded as the corresponding price of $F$.

In this paper, we propose numerical methods of LRM strategies $\xi^F$ and MVH strategies $\vt^F$ for call options when the asset price process is given by an exponential NIG process,
by extending results of \cite{AIS} and \cite{AI}.
Our main contributions are as follows:
\begin{enumerate}
 \item To ensure the existence of LRM and MVH strategies, we need to impose some integrability conditions (Assumption 1.1 of \cite{AIS}) with respect to the L\'evy measure of
       the logarithm of the asset price process.
       Thus, we shall give a sufficient condition in terms of the parameters of NIG processes as our standing assumptions,
       which enables us to check if a parameter set estimated by financial market data satisfies Assumption 1.1 of \cite{AIS}.
 \item The so-called minimal martingale measure (MMM) is indispensable to discuss the LRM problem.
       In particular, the characteristic function of the asset price process under the MMM is needed in the numerical method developed by \cite{AIS}.
       Thus, we provide its explicit representation for NIG models.
 \item In general, a Fourier transform is given as an integration on $[0,\infty)$.
       In fact, we represent LRM strategies by such an improper integration, and truncate its integration interval in order to use FFTs.
       Thus, we shall estimate a sufficient length of the integration interval to reduce the associated truncation error within given allowable extent.
\end{enumerate}
Actually, we need to overcome some complicated calculations in order to achieve the three objects above,
since the L\'evy measure of an NIG process includes a modified Bessel function of the second kind with parameter $1$.

An outline of this paper is as follows:
A precise model description is given in Section 2.
Main results will be stated in Section 3.
Our standing assumption described in terms of the parameters of NIG models is introduced in Subsection 3.1,
which is followed by subsections discussing the characteristic function under the MMM, a representation of LRM strategies, an estimation of the integration interval,
and a representation of MVH strategies.
Note that proofs are postponed until Appendix.
Section 4 is devoted to numerical results.

%
%
\setcounter{equation}{0}
\section{Model description}
We consider throughout a financial market composed of one risk-free asset and one risky asset with finite time horizon $T>0$.
For simplicity, we assume that market's interest rate is zero, that is, the price of the risk-free asset is $1$ at all times.
$(\Omega, \calF, \bbP)$ denotes the canonical L\'evy space, which is given as the product space of spaces of compound Poisson processes on $[0,T]$.
Denote by $\bbF=\{\calF_t\}_{t\in[0,T]}$ the canonical filtration completed for $\bbP$.
For more details on the canonical L\'evy space, see Section 4 of Sol\'e et al. \cite{S07} or Section 3 of Delong and Imkeller \cite{DI}.
Let $L=\{L_t\}_{t\in[0,T]}$ be a pure jump L\'evy process with L\'evy measure $\nu$ defined on $(\Omega, \calF, \bbP)$.
We define the jump measure of $L$ as
\[
N([0,t],A):=\sum_{0\leq u\leq t}{\bf 1}_A(\Delta L_u)
\]
for any $A\in\calB(\bbR_0)$ and any $t\in[0,T]$, where $\Delta L_t:=L_t-L_{t-}$, $\bbR_0:=\bbR\setminus\{0\}$, and $\calB(\bbR_0)$ denotes the Borel $\sigma$-algebra on $\bbR_0$.
In addition, its compensated version $\tN$ is defined as
\[
\tN([0,t],A):=N([0,t],A)-t\nu(A).
\]

In this paper, we study the case where $L$ is given as a normal inverse Gaussian (NIG) process.
Here a pure jump L\'evy process $L$ is called an NIG process with parameters $\alpha>0$, $-\alpha<\beta<\alpha$, $\delta>0$, if its characteristic function is given as
\[
\bbE[e^{izL_t}]=\exp\l\{-\delta\l(\sqrt{\alpha^2-(\beta+iz)^2}-\sqrt{\alpha^2-\beta^2}\r)\r\}
\]
for any $z\in\bbC$ and any $t\in[0,T]$.
Note that the corresponding L\'evy measure $\nu$ is given as
\[
\nu(dx)=\frac{\delta\alpha}{\pi}\frac{e^{\beta x}K_1(\alpha|x|)}{|x|}dx
\]
for $x\in\bbR_0$, where $K_1$ is the modified Bessel function of the second kind with parameter $1$.
When we need to emphasize the model parameters, $\nu$ is denoted by $\nu[\alpha,\beta,\delta]$.
In addition, the process $L$ can also be described as the following time-changed Brownian motion with IG subordinator:
\[
L_t=\beta{\delta^2}I_t+{\delta}B_{I_t},
\]
where $B=\{B_t\}_{t\in[0,T]}$ is a one-dimensional standard Brownian motion, and $I=\{I_t\}_{t\in[0,T]}$ is an IG process with parameter $(1,\delta\sqrt{\alpha^2-\beta^2})$.
For more details on NIG processes, see Section 4.4 of Cont and Tankov \cite{CT} and Subsection 5.3.8 of Schoutens \cite{Scho}.
In this paper, the risky asset price process $S=\{S_t\}_{t\in[0,T]}$ is given as the exponential of the NIG process $L$:
\[
S_t=S_0e^{L_t},
\]
where $S_0>0$.

Now, we prepare some additional notation.
For $v\in[0, \infty)$ and $a\in(\frac{3}{2},2]$, we define
\[
M_1(v,a):=\frac{v^2+\alpha^2-(a+\beta)^2}{\alpha^2}, \hspace{3mm} M_2:=1-\frac{\beta^2}{\alpha^2}, \hspace{3mm} b(v,a):=\frac{2(a+\beta)v}{\alpha^2},
\]
and
\begin{equation}\label{eq-W}
W(v,a):=\frac{\delta\alpha}{\sqrt{2}}\l(i\sqrt{\sqrt{M_1^2+b^2}-M_1}-\sqrt{\sqrt{M_1^2+b^2}+M_1}+\sqrt{2M_2}\r),
\end{equation}
where $M_1(v,a)$ and $b(v,a)$ are abbreviated to $M_1$ and $b$, respectively.
Note that we can define $W(0,1)$ and $W(v,a+1)$ for $v\in[0, \infty)$ and $a\in(\frac{3}{2},2]$ as well.
Moreover, when it is desirable to emphasize the parameters $\alpha,\beta$ and $\delta$, we denote the above four functions as
$M_1(v,a;\alpha,\beta)$, $M_2(\alpha,\beta)$, $b(v,a;\alpha,\beta)$ and $W(v,a;\alpha,\beta,\delta)$, respectively.

%
%
\setcounter{equation}{0}
\section{Main results}
\subsection{Standing assumption}
We introduce our standing assumption in terms of model parameters.

\begin{ass}\label{ass0}
\[
\alpha>\frac{5}{2}, \hspace{3mm} -\frac{3}{2}<\beta\le-\frac{1}{2}, \ \ \ \mbox{and} \ \ \ \beta+4<\alpha.
\]
\end{ass}

\noindent
Now, we show that Assumption \ref{ass0} is a sufficient condition for Assumption 1.1 of \cite{AIS}, which ensures the existence of LRM and MVH strategies.

\begin{prop}\label{prop1}
Under Assumption \ref{ass0}, we have
\begin{enumerate}
\item $\int_{\bbR_0}(e^x-1)^4\nu(dx)<\infty$,
\item $0\geq \int_{\bbR_0}(e^x-1)\nu(dx)> -\int_{\bbR_0}(e^x-1)^2\nu(dx)$.
\end{enumerate}
\end{prop}

\noindent
We postpone the proof of Proposition \ref{prop1} until Appendix \ref{A-cond1}.
Remark that Condition 2 in Proposition \ref{prop1} is the same as the second condition of Assumption 1.1 of \cite{AIS}.
On the other hand, Condition 1 is a modification of the first condition of Assumption 1.1 of \cite{AIS}, which is given as follows:
\[
\int_{\bbR_0}(|x|\vee x^2)\nu(dx)<\infty \mbox{ and } \int_{\bbR_0}(e^x-1)^n\nu(dx)<\infty \mbox{ for }n=2,4.
\]
Firstly, $\int_{\bbR_0}x^2\nu(dx)<\infty$ and $\int_{\bbR_0}(e^x-1)^2\nu(dx)<\infty$ are redundant, since $\int_{\bbR_0}(x^2\wedge1)\nu(dx)<\infty$ holds.
Next, NIG processes do not have the finiteness of $\int_{\bbR_0}|x|\nu(dx)$, different from VG processes.
Actually, $S$ is described by a stochastic integration with respect to $N$ in \cite{AIS}.
Thus, the condition $\int_{\bbR_0}|x|\nu(dx)<\infty$ is needed.
On the other hand, describing $S$ as
\[
S_t = S_0e^{L_t} = S_0\exp\l\{\int_0^t\int_{\bbR_0}x\tN(du,dx)+t\int_{\bbR_0}x\nu(dx)\r\},
\]
we do not need to assume it.

\subsection{The minimal martingale measure}
In this subsection, we focus on the minimal martingale measure (MMM):
an equivalent martingale measure under which any square-integrable $\bbP$-martingale orthogonal to the martingale part of $S$ remains a martingale.
Remark that the MMM plays a vital role in quadratic hedging problems.
Denote $\mu^S:=\int_{\bbR_0}(e^x-1)\nu(dx)$, $C_\nu:=\int_{\bbR_0}(e^x-1)^2\nu(dx)$, $h:=\mu^S/C_\nu$,
and
\[
\theta_x:=\frac{\mu^S(e^x-1)}{C_\nu}
\]
for $x\in\bbR_0$.
As discussed in \cite{AIS}, the MMM $\tP$ exists under Assumption 1.1 of \cite{AIS}, and its Radon-Nikodym density is given as
\[
\frac{d\tP}{d\bbP}=\exp\l\{\int_{\bbR_0}\log(1-\theta_x)\tN([0,T],dx)+T\int_{\bbR_0}\l(\log(1-\theta_x)+\theta_x\r)\nu(dx)\r\}.
\]
Note that $\theta_x<1$ holds for any $x\in\bbR_0$ under Assumption \ref{ass0} by Proposition \ref{prop1}.
Furthermore, $\tP$ is not only the MMM, but also the variance-optimal martingale measure (VOMM) in our setting as discussed in \cite{AI}.
Note that the VOMM is an equivalent martingale measure whose density minimizes the $L^2(\bbP)$-norm among all equivalent martingale measures.
Since MVH strategies are described using the VOMM, we use $\tP$ to express MVH strategies as well as LRM strategies.

Here we prepare some additional notation.
From the view of the Girsanov theorem,
\[
\tN^{\tP}([0,t],dx):=\tN([0,t],dx)+\theta_x\nu(dx)t
\]
is the compensated jump measure of $L$ under $\tP$.
This means that the L\'evy measure under $\tP$, denoted by $\nu^{\tP}$, is given as
\begin{equation}\label{eq-nu*}
\nu^{\tP}(dx)=(1-\theta_x)\nu(dx).
\end{equation}
$L$ is then rewritten as
\begin{equation}\label{eq-L*}
L_t=\int_{\bbR_0}x\tN^{\tP}([0,t],dx)+\mu^*t,
\end{equation}
where $\mu^*:=\int_{\bbR_0}(x-e^x+1)\nu^{\tP}(dx)$, and the stochastic differential equation for $S$ under $\tP$ is given as $dS_t=S_{t-}\int_{\bbR_0}(e^x-1)\tN^{\tP}(dt,dx)$.

In order to develop FFT-based numerical schemes, we need an explicit representation of the characteristic function of $L$ under $\tP$:
\[
\phi_{T-t}(z):=\bbE_{\tP}[e^{izL_{T-t}}]
\]
for $z\in\bbC$.
Before stating it, we calculate $\nu^{\tP}(dx)$ the L\'evy measure of $L$ under $\tP$.
Recall that $\nu[\alpha,\beta,(1+h)\delta](dx)$ represents the L\'evy measure of an NIG process with parameters $\alpha$, $\beta$ and $(1+h)\delta$.
We provide the proof of the following proposition in Appendix \ref{A-prop-nu*}.

\begin{prop}\label{prop-nu*}
We have
\[
\nu^{\tP}(dx)=\nu[\alpha,\beta,(1+h)\delta](dx)+\nu[\alpha,1+\beta,-h\delta](dx).
\]
\end{prop}

Now, we provide a representation of $\phi$ using the function $W(v,a)$ defined in (\ref{eq-W}).
Remark that $W(v,a;\alpha,1+\beta,\delta)$ is also well-defined, since $M_2(\alpha,\beta+1)>0$ by Assumption \ref{ass0}.
The proof of the following proposition is given in Appendix \ref{A-prop2}.

\begin{prop}\label{prop2}
For any $v\in[0,\infty)$ and any $a\in(\frac{3}{2},2]$, we have
\begin{align*}
\phi_{T-t}(v-ia)
&= \exp\l\{(T-t)i(v-ia)\l(\mu^*-\frac{(1+h)\delta\beta}{\sqrt{\alpha^2-\beta^2}}+\frac{h\delta(1+\beta)}{\sqrt{\alpha^2-(1+\beta)^2}}\r)\r\} \\
&  \hspace{5mm}\times\exp\l\{(T-t)\bigg(W(v,a;\alpha,\beta,(1+h)\delta)+W(v,a;\alpha,1+\beta,-h\delta)\bigg)\r\}
\end{align*}
where $\mu^*=\int_{\bbR_0}(x-e^x+1)\nu^{\tP}(dx)$.
\end{prop}

\subsection{Local risk-minimization}
In this subsection, we introduce how to compute LRM strategies for call options $(S_T-K)^+$ with strike price $K>0$.
First of all, we give a precise definition of the LRM strategy for claim $F\in L^2(\bbP)$.
The following is based on Theorem 1.6 of Schweizer \cite{Sch3}.

\begin{defn}
\begin{enumerate}
\item A strategy $\vp=(\xi,\eta)$ is said to be an $L^2$-strategy, if $\xi$ satisfies $\bbE\l[\int_0^TS^2_{u-}\xi^2_udu\r]<\infty$,
      and $V(\vp)$ is a right continuous process with $\bbE[V_t^2(\vp)]<\infty$ for every $t\in[0,T]$.
\item An $L^2$-strategy $\vp$ is called the LRM strategy for claim $F$, if $V_T(\vp^F)=F$, and $[C(\vp^F),M]$ is a uniformly integrable martingale,
      where $M=\{M_t\}_{t\in[0,T]}$ is the martingale part of $S$.
\end{enumerate}
\end{defn}

\noindent
Note that all the conditions of Theorem 1.6 of \cite{Sch3} hold under Assumption 1.1 of \cite{AIS} as seen in Example 2.8 of \cite{AS}.
The above definition of LRM strategies is a simplified version, since the original one, introduced in \cite{Sch} and \cite{Sch3}, is rather complicated.
Now, an $F\in L^2(\bbP)$ admits a F\"ollmer-Schweizer decomposition, if it can be described by
\[
F=F_0+G_T(\xi^{FS})+L_T^{FS},
\]
where $F_0\in\bbR$, $\xi^{FS}=\{\xi^{FS}_t\}_{t\in[0,T]}$ is a predictable process satisfying $\bbE\l[\int_0^TS^2_{u-}(\xi^{FS}_u)^2du\r]<\infty$,
and $L^{FS}=\{L^{FS}_t\}_{t\in[0,T]}$ is a square-integrable martingale orthogonal to $M$ with $L_0^{FS}=0$.
In addition, Proposition 5.2 of \cite{Sch3} provides that, under Assumption 1.1 of \cite{AIS}, the LRM strategy $\vp^F=(\xi^F,\eta^F)$ for $F\in L^2(\bbP)$ exists
if and only if $F$ admits a F\"ollmer-Schweizer decomposition; and its relationship is given by
\[
\xi^F_t=\xi^{FS}_t,\hspace{3mm}\eta^F_t=F_0+G_t(\xi^F)+L^{FS}_t-\xi^F_tS_t.
\]
As a result, it suffices to obtain a representation of $\xi^F$ in order to get $\vp^F$.
Henceforth, we identify $\xi^F$ with $\vp^F$.

We consider call options $(S_T-K)^+$ with strike price $K>0$ as claims to hedge.
Now, we denote $F(K)=(S_T-K)^+$ for $K>0$, and define a function
\[
I(s,t,K):=\int_{\bbR_0}\bbE_{\tP}[(S_Te^x-K)^+-(S_T-K)^+ | S_{t-}=s](e^x-1)\nu(dx).
\]
for $s>0$, $t\in[0,T]$ and $K>0$.
\cite{AS} gave an explicit representation of $\xi^{F(K)}_t$ for any $t\in[0,T]$ and any $K>0$ using Malliavin calculus for L\'evy processes.

\begin{prop}[Proposition 4.6 of \cite{AS}]
For any $K>0$ and any $t\in[0,T]$, 
\begin{align}\label{eq-prop-AS}
\xi^{F(K)}_t=\frac{I(S_{t-},t,K)}{S_{t-}C_\nu}.
\end{align}
\end{prop}

\noindent
In addition, \cite{AIS} introduced an integral representation for $I(S_{t-},t,K)$ as
\[
I(S_{t-},t,K) = \frac{1}{\pi}\int_0^\infty K^{-iv-a+1}\int_{\bbR_0}(e^{(iv+a)x}-1)(e^x-1)\nu(dx)\frac{\phi_{T-t}(v-ia)S_{t-}^{iv+a}}{(iv+a)(iv+a-1)}dv,
\]
where $a\in(1,2]$ and the right-hand side is independent of the choice of $a$.
Remark that we narrow the range of $a$ to $(\frac{3}{2},2]$ for technical reasons, but this does not restrict our development of numerical schemes,
since we take 1.75 as the value of $a$ in our numerical experiments.
To compute $I(S_{t-},t,K)$, we need to calculate the integration $\int_{\bbR_0}(e^{(iv+a)x}-1)(e^x-1)\nu(dx)$.
Now, Lemma \ref{lem5} implies that
\begin{align*}
\int_{\bbR_0}e^{(iv+a)x}(e^x-1)\nu(dx)
&= \int_{\bbR_0}e^{(iv+a)x}(e^x-1)\nu(dx) =\int_{\bbR_0}(e^{(iv+a+1)x}-e^{(iv+a)x})\nu(dx) \\
&= \int_{\bbR_0}(e^{(iv+a+1)x}-1)\nu(dx)-\int_{\bbR_0}(e^{(iv+a)x}-1)\nu(dx) \\
&= W(v,a+1)-W(v,a),
\end{align*}
from which we have
\begin{equation}\label{eq-I}
I(S_{t-},t,K) = \frac{1}{\pi}\int_0^\infty K^{-iv-a+1}\Big(W(v,a+1)-W(v,a)-W(0,1)\Big)\frac{\phi_{T-t}(v-ia)S_{t-}^{iv+a}}{(iv+a)(iv+a-1)}dv.
\end{equation}
Thus, we can compute $I(S_{t-},t,K)$ using the FFT as mentioned in \cite{AIS}.

\subsection{Integration interval}
To compute the integral (\ref{eq-I}) with the FFT, we discretize $I(S_{t-},t,K)$ as
\[
I(S_{t-},t,K) \approx \frac{1}{\pi}\sum^{N-1}_{j=0}e^{(-i\eta j-a+1)\log K}\bigg(W(\eta j,a+1)-W(\eta j,a)-W(0,1)\bigg)
                      \frac{\phi_{T-t}(\eta j-ia)S_{t-}^{i\eta j+a}}{(i\eta j+a)(i\eta j+a-1)}\eta,
\]
where $N$ represents the number of grid points, and $\eta>0$ is the distance between adjacent grid points.
This approximation corresponds to the integral (\ref{eq-I}) over the interval $[0,N\eta]$, so we need to specify $N$ and $\eta$ to satisfy
\begin{equation}\label{eq-CM3}
\l|\frac{1}{\pi}\int_{N\eta}^\infty K^{-iv-a+1}\Big(W(v,a+1)-W(v,a)-W(0,1)\Big)\frac{\phi_{T-t}(v-ia)S_{t-}^{iv+a}}{(iv+a)(iv+a-1)}dv\r|<\ve
\end{equation}
for a given sufficiently small value $\ve>0$, which represents the allowable error.
Thus, we shall estimate a sufficient length for the integration interval of (\ref{eq-I}) for a given allowable error $\ve>0$ in the sense of (\ref{eq-CM3}).
The following proposition is shown in Appendix \ref{A-prop3}

\begin{prop}\label{prop3}
For $\ve>0$ and $t\in[0,T)$, if $w>1$ satisfies
\begin{equation}\label{eq-interval}
\frac{\sqrt{2}K^{-a+1}S^a_{t-}C(t)}{\pi{(T-t)}\ve}\l(2+\sqrt{\alpha^2-(a+\beta)^2+2(a+1+\beta)^2}\r) < e^{(T-t)\delta w},
\end{equation}
we have
\[
\l|\frac{1}{\pi}\int_w^\infty K^{-iv-a+1}\Big(W(v,a+1)-W(v,a)-W(0,1)\Big)\frac{\phi_{T-t}(v-ia)S_{t-}^{iv+a}}{(iv+a)(iv+a-1)}dv\r|<\ve,
\]
where $C(t)$ is defined as
\begin{align}
C(t) & :=\exp\l\{(T-t)a\l(\mu^*-\frac{(1+h)\delta\beta}{\sqrt{\alpha^2-\beta^2}}+\frac{h\delta(1+\beta)}{\sqrt{\alpha^2-(1+\beta)^2}}\r)\r\} \notag \\
     & \hspace{5mm}\times\exp\l\{(T-t)\delta\alpha\l((1+h)\sqrt{M_2(\alpha,\beta)}-h\sqrt{M_2(\alpha,1+\beta)}\r)\r\}
\label{eq-prop3}
\end{align}
for any $t\in[0,T)$.
\end{prop}

\begin{rem}
In Proposition \ref{prop3}, the case of $t=T$ is excluded, but this does not restrict our numerical method,
since we do not need to compute the value of LRM strategies when the time to maturity $T-t$ is 0.
\end{rem}

\subsection{Mean-variance hedging}
As introduced in Section 1, the MVH strategy for claim $F\in L^2(\bbP)$ is defined as the solution $(\vt^F,c^F)$ to the minimization problem
\[
\min_{c\in\bbR, \vt\in\Theta}\bbE\l[\big(F-c-G_T(\vt)\big)^2\r],
\]
where $\Theta$ is the set of all admissible strategies,
mathematically the set of $\bbR$-valued $S$-integrable predictable processes $\vt$ satisfying $\bbE\l[\int_0^T\vt_u^2S_{u-}^2du\r]<\infty$.
\cite{AI} gave an explicit closed-form representation of $\vt^F$ for exponential additive models,
and developed a numerical scheme for call options $(S_T-K)^+$ with strike price $K>0$ for exponential L\'evy models.
Different from LRM strategies, the value of $\vt^F_t$ is depending on not only $S_{t-}$, but also the whole trajectory of $S$ from $0$ to $t-$.
However it is impossible to observe the trajectory of $S$ continuously.
Thus, \cite{AI} developed a numerical scheme to compute $\vt^F_t$ approximately using discrete observational data $S_{t_0}, S_{t_1},\dots,S_{t_n}$,
where $n\geq1$ and $t_k:=\frac{kt}{n+1}$.

We need some preparations before introducing the representation of $\vt^F_t$ obtained by \cite{AI}.
Firstly, we consider the VOMM, which is an equivalent martingale measure whose density minimizes the $L^2(\bbP)$-norm among all equivalent martingale measures.
Indeed, the MMM $\tP$ coincides with the VOMM in our setting as mentioned in Subsection 3.2.
Next, we define a process $\calE=\{\calE_t\}_{t\in[0,T]}$ as a solution to the stochastic differential equation $\calE_t=1-h\int_0^t\calE_{u-}dS_u$,
and $H^F=\{H^F_t\}_{t\in[0,T]}$ as $H^F_t:=\bbE_{\tP}[F | S_{t-}]$.
Moreover, remark that Assumption 2.1 of \cite{AI} is satisfied under Assumption \ref{ass0}.

From the view of \cite{AI}, the MVH strategy $\vt^{F(K)}_t$ for call option $F(K)=(S_T-K)^+$ is represented in closed-form as
\[
\vt^{F(K)}_t = \xi^{F(K)}_t+\frac{h\calE_{t-}}{S_{t-}}\int_0^{t-}\frac{dH^{F(K)}_u-\xi^{F(K)}_udS_u}{\calE_u}.
\]
Now, the process $H^{F(K)}_t=\bbE_{\tP}[F(K)|S_{t-}]$ is represented as
\[
H^{F(K)}_t = \frac{1}{\pi}\int_0^\infty K^{-iv-a+1}\frac{\phi_{T-t}(v-ia)S_{t-}^{iv+a}}{(iv+a-1)(iv+a)}dv,
\]
which is computable with the FFT.
As a result, using discrete observational data $S_{t_0}, S_{t_1},\dots,S_{t_n}$, we can approximate $\vt^{F(K)}_t$ as
\begin{equation}\label{eq-approx}
\vt^{F(K)}_t\approx\xi^{F(K)}_t+\frac{h\calE_{t_n}}{S_{t_n}}\sum_{k=1}^n\frac{\Delta H^{F(K)}_{t_k}-\xi^{F(K)}_{t_k}\Delta S_{t_k}}{\calE_{t_k}},
\end{equation}
where $H^{F(K)}_{t_k}=\bbE_{\tP}[F(K)|S_{t_k}]$ and $t_k:=\frac{kt}{n+1}$ for $k=0,1,\dots,n$; $t$ is corresponding to $t_{n+1}$; and, for $k=1,\dots,n$,
we denote $\Delta X_{t_k}:=X_{t_k}-X_{t_{k-1}}$ for a process $X$ and
\[
\calE_{t_{k+1}}=\calE_{t_k}\l\{1-\frac{h\Delta S_{t_{k+1}}}{S_{t_k}}\r\}
\]
with $\calE_{t_0}=1$.

%
%
\setcounter{equation}{0}
\section{Numerical results}
\begin{figure}[htbp]
\begin{center}
   \includegraphics[width=100mm]{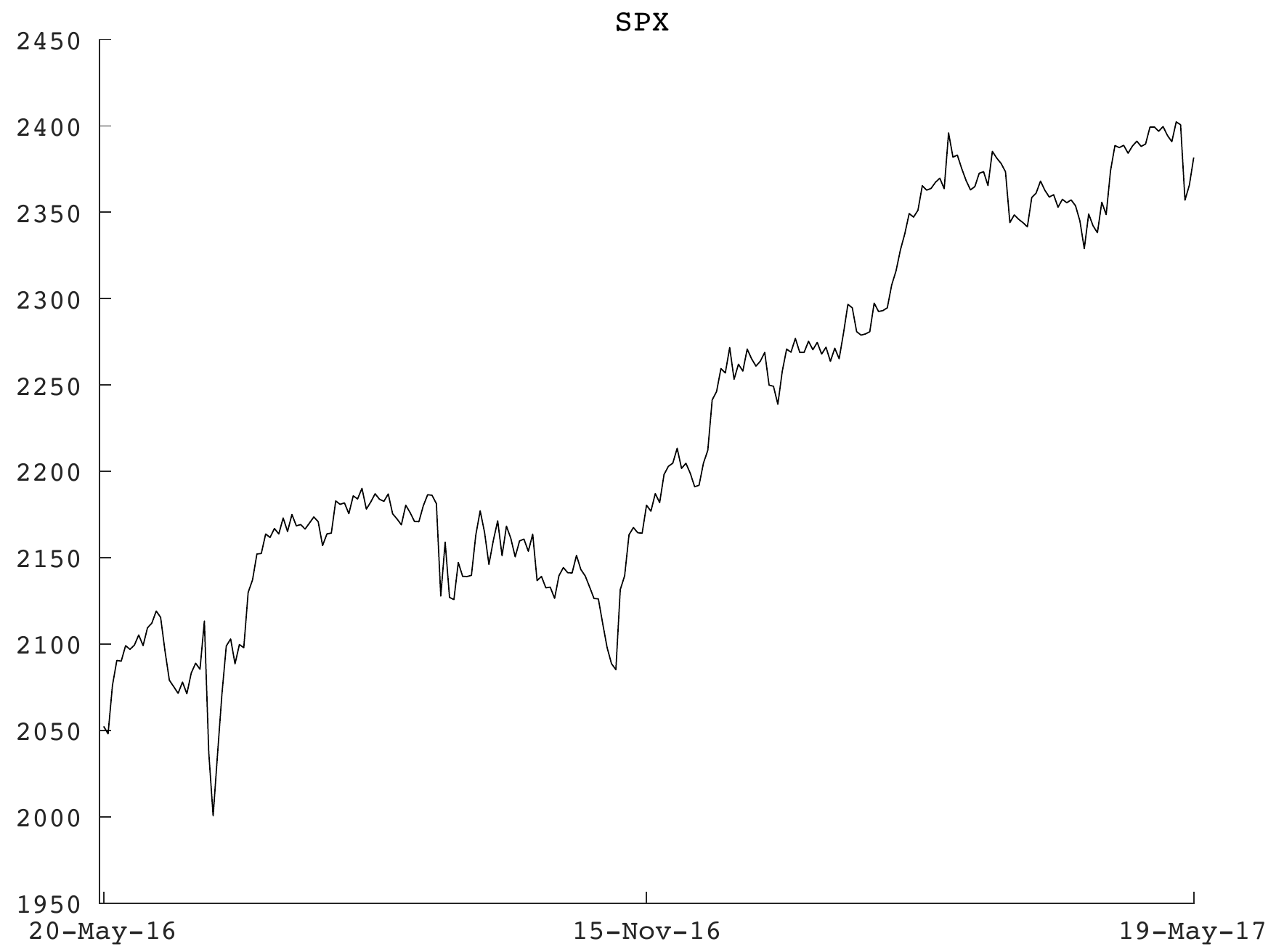}
\end{center}
\caption{SPX dairy closing prices.}
\label{fig1}
\end{figure}

We consider European call options on the S\&P 500 Index (SPX) matured on 19 May 2017, and set the initial date of our hedging to 20 May 2016.
We fix $T$ to $1$.
There are 250 business days on and after 20 May 2016 until and including 19 May 2017.
For example, 20 May 2016 and 23 May 2016 are corresponding to time $0$ and $\frac{1}{249}$, respectively, since 20 May 2016 is Friday.
Note that we shall use 250 dairy closing prices of the SPX on and after 20 May 2016 until and including 19 May 2017 as discrete observational data.
Figure \ref{fig1} illustrates the fluctuation of the SPX.

Next, we set model parameters as
\[
\l\{
\begin{array}{l}
\alpha= 25.61598030765035, \\
\beta = -1.2668546614155765, \\
\delta= 0.40532772478162127,
\end{array}\r.
\]
which are calibrated by the data set of European call options on the SPX at 20 April 2016.
Note that the above parameter set satisfies Assumption \ref{ass0}.
Moreover, we choose
\[
N=2^{16}, \eta=0.25 \ \mbox{and} \ a=1.75
\]
as parameters related to the FFT, that is, $N\eta=2^{14}$, which satisfies (\ref{eq-interval}) for any $t\leq\frac{248}{249}$ when we take $\ve=0.01$ as our allowable error.

\begin{figure}[htbp]
\begin{center}
   \includegraphics[width=100mm]{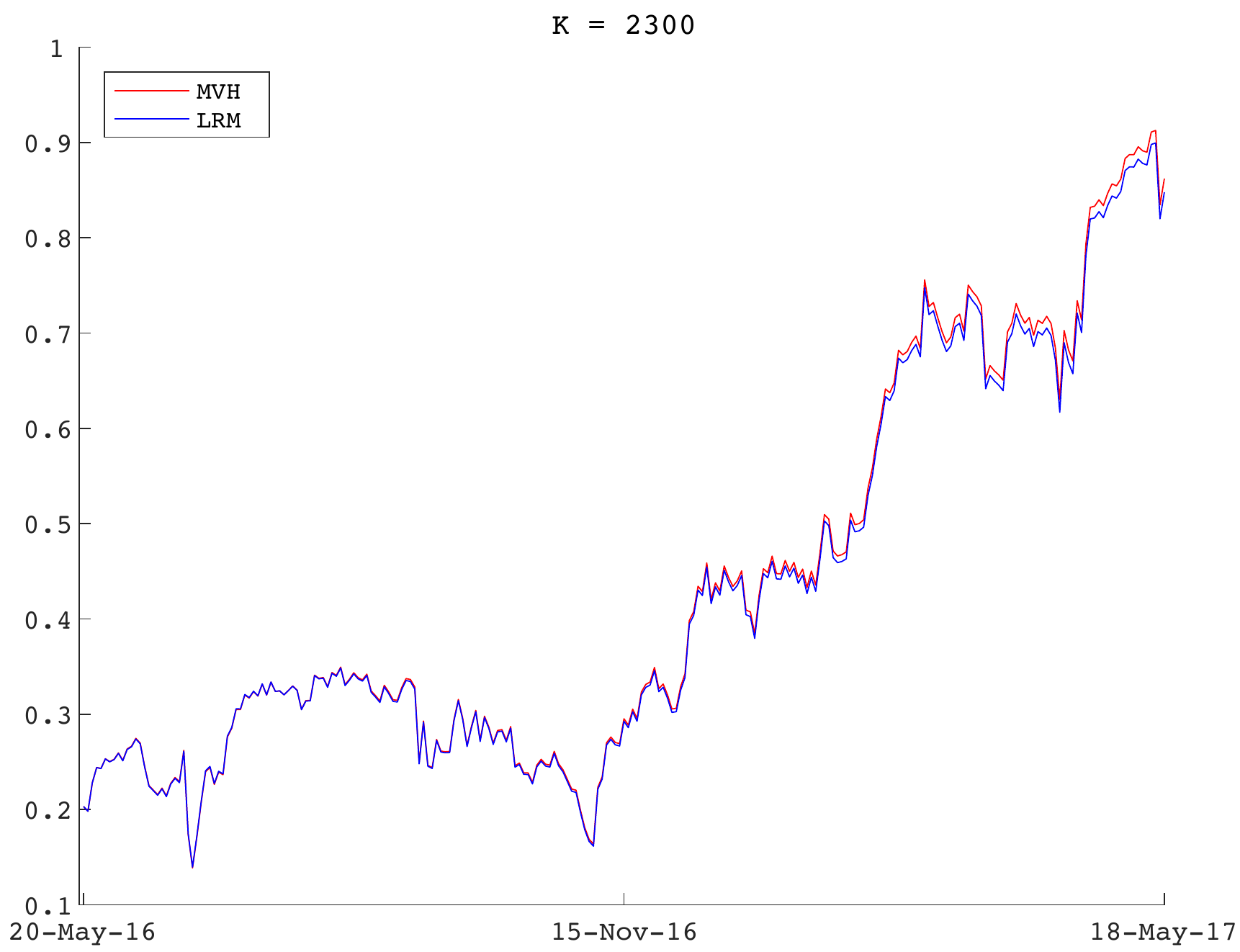}
\end{center}
\caption{Values of LRM strategies $\xi^{F(K)}_t$ and MVH strategies $\vt^{F(K)}_t$ for $K=2300$.
         The blue and the red lines represent the values of $\xi^{F(K)}_t$ and $\vt^{F(K)}_t$, respectively.
         The two lines are almost overlapping when $t$ is small; and separate gradually as drawing near to the maturity.}
\label{fig2}
\end{figure}

\begin{figure}[htbp]
\begin{center}
   \includegraphics[width=100mm]{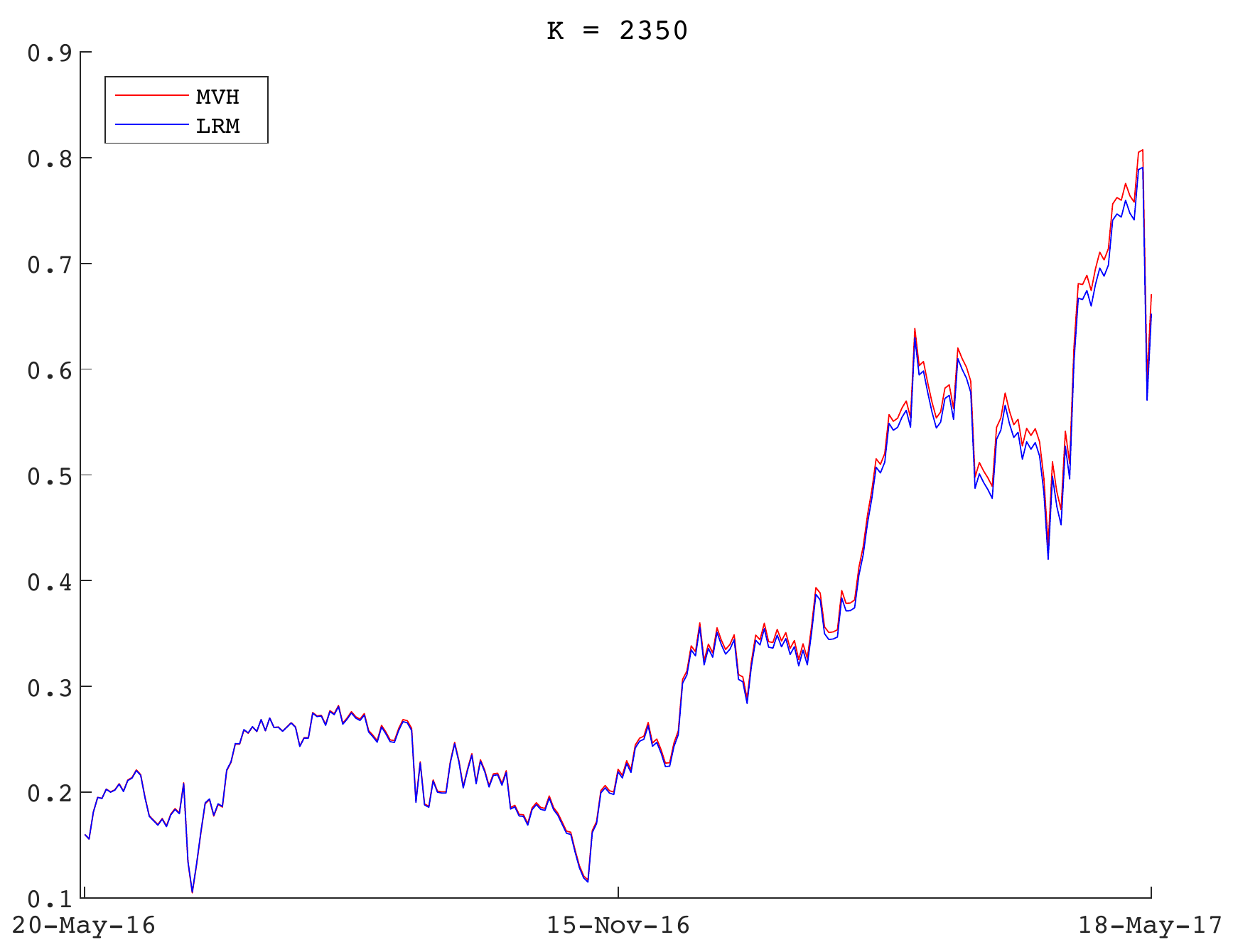}
\end{center}
\caption{Values of LRM strategies $\xi^{F(K)}_t$ and MVH strategies $\vt^{F(K)}_t$ for $K=2350$.}
\label{fig3}
\end{figure}

\begin{figure}[htbp]
\begin{center}
   \includegraphics[width=100mm]{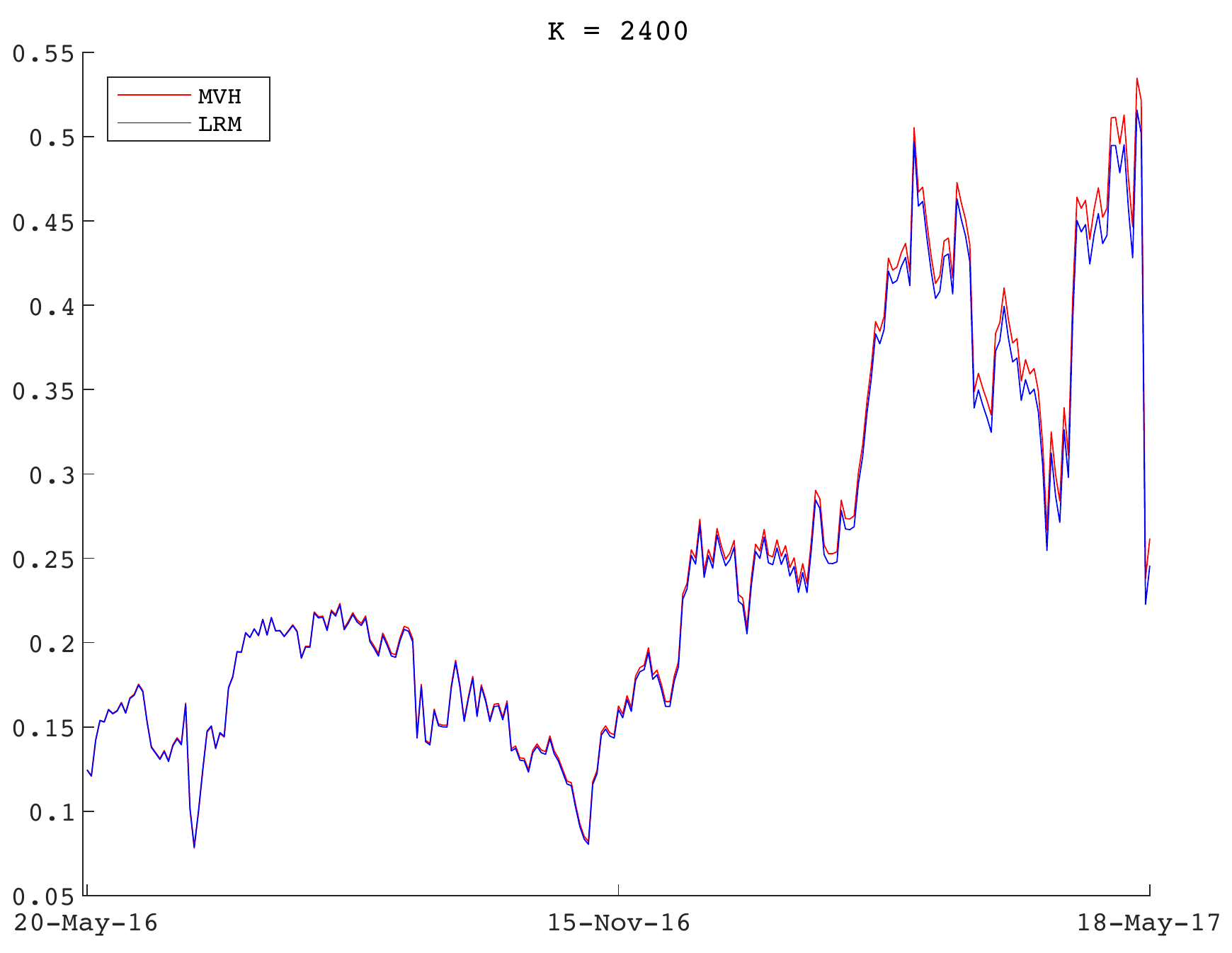}
\end{center}
\caption{Values of LRM strategies $\xi^{F(K)}_t$ and MVH strategies $\vt^{F(K)}_t$ for $K=2400$.}
\label{fig4}
\end{figure}

As contingent claims to hedge, we consider call options with strike price $K=$2300, 2350 and 2400; and
compute the values of LRM strategies $\xi^{F(K)}_t$ and MVH strategies $\vt^{F(K)}_t$ for $t=\frac{1}{249}, \frac{2}{249},\dots,1$
by using (\ref{eq-prop-AS}), (\ref{eq-I}) and (\ref{eq-approx}).
Remark that, for $k=1,\dots,249$, $\xi^{F(K)}_{\frac{k}{249}}$ and $\vt^{F(K)}_{\frac{k}{249}}$ are constructed on time $\frac{k-1}{249}$
using observational data $S_0,S_{\frac{1}{249}},\dots,S_{\frac{k-1}{249}}$.
Figures \ref{fig2} -- \ref{fig4} show the values of $\xi^{F(K)}_t$ and $\vt^{F(K)}_t$ versus times $t=\frac{1}{249},\frac{2}{249},\dots,1$
for the case where $K=2300, 2350$ and 2400, respectively.

\break

%
%
\appendix
\setcounter{equation}{0}
\section{Appendix}
\renewcommand{\theequation}{A.\arabic{equation}}
\setcounter{equation}{0}
\subsection{Proof of Proposition \ref{prop1}}\label{A-cond1}
In order to see Condition 1, it suffices to show $\int_1^\infty(e^x-1)^4\nu(dx)<\infty$ and $\int^{-1}_{-\infty}(e^x-1)^4\nu(dx)<\infty$.

Firstly, we see $\int_1^\infty(e^x-1)^4\nu(dx)<\infty$.
Noting that the Sommerfeld integral representation for the function $K_1$ (see, e.g. Appendix A of \cite{CT}):
\begin{align}\label{eq-Sommerfeld}
K_1(z)=\frac{z}{4}\int_0^\infty\exp\l\{-s-\frac{z^2}{4s}\r\}s^{-2}ds
\end{align}
for $z\geq0$, we have
\begin{align*}
\int_1^\infty(e^x-1)^4{\nu}(dx)
&=    \frac{\delta\alpha}{\pi}\int_1^\infty(e^x-1)^4\frac{e^{\beta x}K_1(\alpha x)}{x}dx \\
&=    \frac{\delta\alpha}{\pi}\int_\alpha^\infty(e^\frac{z}{\alpha}-1)^4\exp\l\{\frac{\beta}{\alpha}z\r\}\frac{1}{z}\int_0^\infty\frac{z}{4}\exp\l\{-s-\frac{z^2}{4s}\r\}s^{-2}dsdz \\
&\leq \frac{\delta\alpha}{4\pi}\int_\alpha^\infty\exp\l\{\frac{4+\beta}{\alpha}z\r\}\int_0^\infty\frac{z}{\alpha}\exp\l\{-s-\frac{z^2}{4s}\r\}s^{-2}dsdz \\
&=    \frac{\delta}{4\pi}\int_0^\infty e^{-s}s^{-2}\int_\alpha^\infty\frac{z}{\sqrt{2\pi2s}}\exp\l\{-\frac{1}{4s}\l(z-2s\frac{4+\beta}{\alpha}\r)^2\r\}dz \\
&     \hspace{5mm}\times\exp\l\{\l(\frac{4+\beta}{\alpha}\r)^2s\r\}{\sqrt{2\pi2s}}ds \\
&\leq \frac{\delta}{4\pi}\int_0^\infty e^{-s}s^{-2}\cdot2s\frac{4+\beta}{\alpha}\cdot\exp\l\{\l(\frac{4+\beta}{\alpha}\r)^2s\r\}{\sqrt{2\pi2s}}dt \\
&=    \frac{\delta}{\sqrt{\pi}}\frac{4+\beta}{\alpha}\int_0^\infty s^{-\frac{1}{2}}\exp\l\{\l(\l(\frac{4+\beta}{\alpha}\r)^2-1\r)s\r\}ds \\
&=    \delta(4+\beta)\l(\alpha^2-(4+\beta)^2\r)^{-\frac{1}{2}} < \infty.
\end{align*}
Remark that the above first inequality is given from $(e^\frac{z}{\alpha}-1)^4 \leq e^\frac{4z}{\alpha}$ for any $z\in[\alpha,\infty)$.

Next, we show $\int^{-1}_{-\infty}(e^x-1)^4\nu(dx)<\infty$ by a similar argument to the above.
Noting that $(e^{\frac{z}{\alpha}}-1)^4\leq1$ for any $z\in(-\infty,-\alpha]$, we have
\begin{align*}
\int^{-1}_{-\infty}(e^x-1)^4\nu(dx)
&\leq \frac{\delta\alpha}{4\pi}\int^{-\alpha}_{-\infty}(e^\frac{z}{\alpha}-1)^4
      \exp\l\{\frac{\beta}{\alpha}z\r\}\int_0^\infty\frac{z}{\alpha}\exp\l\{-s-\frac{z^2}{4s}\r\}s^{-2}dsdz \\
&\leq \frac{\delta}{4\pi}\int_0^\infty e^{-s}s^{-2}\int_{-\infty}^\infty\frac{z}{\sqrt{2\pi2s}}\exp\l\{-\frac{1}{4s}\l(z-2s\frac{\beta}{\alpha}\r)^2\r\}dz
      \exp\l\{\frac{\beta^2}{\alpha^2}s\r\}\sqrt{2\pi2s}ds \\
&<    \infty.
\end{align*}
Thus, Condition 1 holds true.

To confirm Condition 2, we need some preparations.
The following lemma is proven in Appendix \ref{A-lem5}

\begin{lem}\label{lem5}
For any $v\in[0,\infty)$ and any $a\in(\frac{3}{2},2]$, we have
\begin{equation}\label{eq-lem5}
\int_{\bbR_0}\l(e^{(iv+a)x}-1\r)\nu(dx)=W(v,a).
\end{equation}
In addition, (\ref{eq-lem5}) still holds for the case where $(v,a)=(0,1)$ and $(v,a+1)$.
\end{lem}

\noindent
We have $\int_{\bbR_0}(e^x-1)\nu(dx)=W(0,1)={\delta\alpha}(\sqrt{M_2}-\sqrt{M_1(0,1)})$.
Assumption \ref{ass0} implies
\[
M_2-M_1(0,1) = \frac{1}{\alpha^2}\l((\alpha^2-\beta^2)-(\alpha^2-(1+\beta)^2)\r) = \frac{1+2\beta}{\alpha^2} \leq 0,
\]
from which the inequality $0 \geq \int_{\bbR_0}(e^x-1)\nu(dx)$ holds true.
To see the second inequality, since we have
\[
-\int_{\bbR_0}(e^x-1)^2\nu(dx) = -\int_{\bbR_0}\l((e^{2x}-1)-2(e^x-1)\r)\nu(dx) = -W(0,2)+2W(0,1),
\]
it suffices to show $W(0,2)-W(0,1)>0$.
Firstly, we have
\begin{align*}
W(0,2)-W(0,1) &= \frac{\delta\alpha}{\sqrt{2}}\l(\l(-\sqrt{2M_1(0,2)}+\sqrt{2M_2}\r)-\l(-\sqrt{2M_1(0,1)}+\sqrt{2M_2}\r)\r) \\
              &= \delta\alpha\l(-\sqrt{M_1(0,2)}+\sqrt{M_1(0,1)}\r).
\end{align*}
On the other hand, it holds that
\[
M_1(0,1)-M_1(0,2) = \frac{\alpha^2-(1+\beta)^2-\alpha^2+(2+\beta)^2}{\alpha^2} = \frac{3+2\beta}{\alpha^2} > 0
\]
by Assumption \ref{ass0}.
As a result, the inequality $\int_{\bbR_0}(e^x-1)\nu(dx) > -\int_{\bbR_0}(e^x-1)^2\nu(dx)$ holds under Assumption \ref{ass0}.
\fin

\subsection{Proof of Lemma \ref{lem5}}\label{A-lem5}
We begin with the following lemma:

\begin{lem}\label{lem-A1}
For any $\gamma\ge0$ and any $M>0$, we have
\[
\int_0^\gamma\int_0^\infty e^{(iu-M)s}s^{-\frac{1}{2}}dsdu = \sqrt{2\pi}\l(\sqrt{\sqrt{M^2+\gamma^2}-M}+i\l(\sqrt{\sqrt{M^2+\gamma^2}+M}-\sqrt{2M}\r)\r)
\]
\end{lem}

\proof
Remark that the characteristic function of the Gamma distribution with parameters $\theta>0$ and $k>0$ is given as
\[
\int_0^\infty e^{iux}\frac{\theta^k}{\Gamma(k)}x^{k-1}e^{-\theta x}dx = \l(\frac{\theta}{\theta-iu}\r)^k
\]
for any $u\in\bbR$, where $\Gamma(\cdot)$ is the Gamma function.
We have then
\[
\int_0^\infty e^{(iu-M)s}s^{-\frac{1}{2}}ds = \sqrt{\frac{M}{M-iu}}\frac{\Gamma\l(\frac{1}{2}\r)}{\sqrt{M}} = \frac{\sqrt{\pi}}{\sqrt{M-iu}}
\]
for any $M>0$ and any $u\in\bbR$.
Thus, we obtain
\[
\int_0^\infty e^{(iu-M)s}s^{-\frac{1}{2}}dt = \sqrt{\frac{\pi}{2}}\l(\frac{\sqrt{\sqrt{M^2+u^2}+M}}{\sqrt{M^2+u^2}}+i\frac{\sqrt{\sqrt{M^2+u^2}-M}}{\sqrt{M^2+u^2}}\r).
\]
Putting $x=\sqrt{M^2+u^2}$, we have
\[
\int_0^\gamma\frac{\sqrt{\sqrt{M^2+u^2}+M}}{\sqrt{M^2+u^2}}du = \int_M^{\sqrt{M^2+\gamma^2}}\frac{\sqrt{x+M}}{\sqrt{x^2-M^2}}dx = 2\sqrt{\sqrt{M^2+\gamma^2}-M}
\]
and
\[
\int_0^\gamma\frac{\sqrt{\sqrt{M^2+u^2}-M}}{\sqrt{M^2+u^2}}du = 2\sqrt{\sqrt{M^2+\gamma^2}+M}-2\sqrt{2M}.
\]
This completes the proof of Lemma \ref{lem-A1}.
\fin

Now, let us go back to the proof of Lemma \ref{lem5}.
For any $v\in[0,\infty)$ and any $a\in(\frac{3}{2},2]$, the same sort of argument as in Appendix \ref{A-cond1} implies that
\begin{equation}\label{eq4:1:10}
\int_{\bbR_0}\l(e^{(iv+a)x}-1\r)\nu(dx)
= \frac{\delta\alpha}{2\sqrt{\pi}}\int_0^\infty e^{-s}s^{-\frac{3}{2}}\int_{\bbR_0}\frac{e^{(iv+a)\frac{z}{\alpha}}-1}{\sqrt{2\pi2s}}
  \exp\l\{-\frac{1}{4s}\l(z-2s\frac{\beta}{\alpha}\r)^2\r\}dz\exp\l\{\frac{\beta^2}{\alpha^2}s\r\}ds.
\end{equation}
Since we have
\[
\int_{\bbR_0}e^{(iv+a)\frac{z}{\alpha}}\exp\l\{-\frac{1}{4s}\l(z-2s\frac{\beta}{\alpha}\r)^2\r\}\frac{dz}{\sqrt{2\pi2s}}
= \exp\l\{i\frac{2s}{\alpha^2}(v\beta+va)-\frac{s}{\alpha^2}(v^2-a^2-2a\beta)\r\},
\]
we obtain
\begin{align}
\mbox{(\ref{eq4:1:10})}
&= \frac{\delta\alpha}{2\sqrt{\pi}}\int_0^\infty\exp\l\{\l(\frac{\beta^2}{\alpha^2}-1\r)s\r\}s^{-\frac{3}{2}}
   \l(\exp\l\{i\frac{2s}{\alpha^2}(v\beta+va)-\frac{s}{\alpha^2}(v^2-a^2-2a\beta)\r\}-1\r)ds \notag \\
&= \frac{\delta\alpha}{2\sqrt{\pi}}\int_0^\infty s^{-\frac{3}{2}}\l(e^{ibs-M_1s}-e^{-M_2s}\r)ds \notag \\
&= \frac{\delta\alpha}{2\sqrt{\pi}}\int_0^\infty s^{-\frac{1}{2}}\l(ie^{-M_1s}\int_0^be^{ius}du+\int_{M_1}^{M_2}e^{-us}du\r)ds \notag \\
&= \frac{\delta\alpha}{2\sqrt{\pi}}i\int_0^b\int_0^\infty e^{(iu-M_1)s}s^{-\frac{1}{2}}dsdu+\frac{\delta\alpha}{2\sqrt{\pi}}\int_{M_1}^{M_2}\int_0^\infty e^{-us}s^{-\frac{1}{2}}dsdu.
\label{eq4:1:11}
\end{align}
On the other hand, we have
\[
\int_{M_1}^{M_2}\int_0^\infty e^{-us}s^{-\frac{1}{2}}dsdu= \int_{M_1}^{M_2}\frac{\sqrt{\pi}}{\sqrt{u}}du = 2\sqrt{\pi}(\sqrt{M_2}-\sqrt{M_1})
\]
by $M_1, M_2>0$.
As a result, using Lemma \ref{lem-A1}, we obtain
\begin{align*}
\mbox{(\ref{eq4:1:11})}
&= \frac{\delta\alpha}{2\sqrt{\pi}}\sqrt{2{\pi}}i\l\{\sqrt{\sqrt{M_1^2+b^2}-M_1}+i\l(\sqrt{\sqrt{M_1^2+b^2}+M_1}-\sqrt{2M_1}\r)\r\} \\
&  \hspace{5mm}+\frac{\delta\alpha}{2\sqrt{\pi}}2\sqrt{\pi}\l(\sqrt{M_2}-\sqrt{M_1}\r)\\
& =\frac{\delta\alpha}{\sqrt{2}}\l\{i\sqrt{\sqrt{M_1^2+b^2}-M_1}-\sqrt{\sqrt{M_1^2+b^2}+M_1}+\sqrt{2M_2}\r\},
\end{align*}
from which (\ref{eq-lem5}) follows for any $v\in[0,\infty)$ and any $a\in(\frac{3}{2},2]$.

For $v\ge0$, we see that (\ref{eq-lem5}) still holds for $a+1$.
To this end, it is enough to make sure that $M_1(v,a+1)$ and $b(v,a+1)$ remain nonnegative.
In fact, we have
\[
M_1(v,a+1) = \frac{\alpha^2-(a+1+\beta)^2}{\alpha^2} \geq 0,
\]
and
\[
b(v,a+1) = \frac{2(a+1+\beta)v}{\alpha^2} \geq 0
\]
by Assumption \ref{ass0}.
Similarly, (\ref{eq-lem5}) follows for the case of $(v,a)=(0,1)$, since
\[
M_1(0,1) = \frac{\alpha^2-(1+\beta)^2}{\alpha^2} \geq \frac{6}{\alpha^2}>0
\]
and $b(0,1)=0$.
\fin

\subsection{Proof of Proposition \ref{prop-nu*}}\label{A-prop-nu*}
Noting that $0\geq h>-1$ by Assumption \ref{ass0} and Proposition \ref{prop1}, we have
\begin{align*}
\nu^{\tP}(dx) &= (1-\theta_x)\nu(dx) = (1-h(e^x-1))\nu(dx) = (1+h)\nu(dx)-he^x\nu(dx) \\
              &= \nu[\alpha,\beta,(1+h)\delta](dx)+\nu[\alpha,1+\beta,-h\delta](dx)
\end{align*}
by (\ref{eq-nu*}).
This completes the proof of Proposition \ref{prop-nu*}.
\fin

\subsection{Proof of Proposition \ref{prop2}}\label{A-prop2}
To show Proposition \ref{prop2}, we start with the following lemma:

\begin{lem}\label{lem-A2}
We have
\[
\int_{\bbR_0}x\nu(dx) = \frac{\delta\beta}{\sqrt{\alpha^2-\beta^2}}.
\]
\end{lem}

\proof
The Sommerfeld integral representation (\ref{eq-Sommerfeld}) implies that
\begin{align*}
\int_{\bbR_0}x\nu(dx)
&= \frac{\delta}{4\pi}\int_{\bbR_0}z\exp\l\{\frac{\beta}{\alpha}z\r\}\int_0^\infty\exp\l\{-s-\frac{z^2}{4s}\r\}s^{-2}dsdz \\
&= \frac{\delta}{4\pi}\int_0^\infty\int_{\bbR_0}\frac{z}{\sqrt{2\pi 2s}}\exp\l\{-\frac{1}{4s}\l(z-2s\frac{\beta}{\alpha}\r)^2\r\}dz
   \sqrt{2\pi 2s}\exp\l\{\frac{\beta^2}{\alpha^2}s\r\}s^{-2}e^{-s}ds \\
&= \frac{\delta\beta}{\sqrt{\pi}\alpha}\int_0^\infty\exp\l\{-\l(1-\frac{\beta^2}{\alpha^2}\r)s\r\}s^{-\frac{1}{2}}ds = \frac{\delta\beta}{\sqrt{\alpha^2-\beta^2}}.
\end{align*}
\fin

\noindent
Note that we do not need Assumption \ref{ass0} in the above proof.
Now, we show Proposition \ref{prop2}.
By  Lemma \ref{lem-A2} and Proposition \ref{prop-nu*}, we have
\[
\int_{\bbR_0}(iv+a)x\nu^{\tP}(dx) = (iv+a)\l(\frac{(1+h)\delta\beta}{\sqrt{\alpha^2-\beta^2}}-\frac{h\delta(1+\beta)}{\sqrt{\alpha^2-(1+\beta)^2}}\r).
\]
Remark that $W(v,a;\alpha,1+\beta,-h\delta)$ is well-defined, and satisfies (\ref{eq-lem5}),
since we have $M_1(v,a;\alpha,\beta+1) = M_1(v,a+1;\alpha,\beta) \geq 0$ and $b(v,a;\alpha,\beta+1) = b(v,a+1;\alpha,\beta) \geq 0$.
(\ref{eq-L*}) implies that
\begin{align*}
\phi_{T-t}(v-ia)
&= \bbE_{\tP}\l[e^{(iv+a)L_{T-t}}\r] \\
&= \bbE_{\tP}\l[\exp\l\{(T-t)(iv+a)\mu^*+\int_{\bbR_0}(iv+a)x\tN^{\tP}([0,T-t],dx)\r\}\r] \\
&= \exp\l\{(T-t)\l((iv+a)\mu^*+\int_{\bbR_0}\l(e^{(iv+a)x}-1-(iv+a)x\r)\nu^{\tP}(dx)\r)\r\} \\
&= \exp\l\{(T-t)(iv+a)\l(\mu^*-\frac{(1+h)\delta\beta}{\sqrt{\alpha^2-\beta^2}}+\frac{h\delta(1+\beta)}{\sqrt{\alpha^2-(1+\beta)^2}}\r)\r\} \\
&  \hspace{5mm}\times\exp\l\{(T-t)\bigg(W(v,a;\alpha,\beta,(1+h)\delta)+W(v,a;\alpha,1+\beta,-h\delta)\bigg)\r\},
\end{align*}
from which Proposition \ref{prop2} follows.
\fin

\subsection{Proof of Proposition \ref{prop3}}\label{A-prop3}
To see Proposition \ref{prop3}, we prepare one proposition and one lemma.
In order to emphasize the parameters $\alpha,\beta$ and $\delta$,
we write $M_1(v,a)$, $M_2$ and $b(v,a)$ as $M_1(v,a;\alpha,\beta)$, $M_2(\alpha,\beta)$ and $b(v,a;\alpha,\beta)$, respectively.

\begin{prop}\label{prop-A1}
For any $v\in[0,\infty)$ and any $t\in[0,T)$, we have
\begin{align*}
|\phi_{T-t}(v-ia)| \leq C(t)e^{-(T-t)\delta v},
\end{align*}
where $C(t)$ is given in (\ref{eq-prop3})
\end{prop}

\proof
Proposition \ref{prop2} implies that
\begin{align*}
&|\phi_{T-t}(v-ia)| \\
&=    \Bigg|\exp\l\{(T-t)(iv+a)\l(\mu^*-\frac{(1+h)\delta\beta}{\sqrt{\alpha^2-\beta^2}}+\frac{h\delta(1+\beta)}{\sqrt{\alpha^2-(1+\beta)^2}}\r)\r\} \\
&     \hspace{5mm}\times\exp\l\{(T-t)\Big(W(v,a;\alpha,\beta,(1+h)\delta)+W(v,a;\alpha,1+\beta,-h\delta)\Big)\r\}\Bigg| \\
&=    C(t)\exp\l\{-(T-t)\frac{(1+h)\delta\alpha}{\sqrt{2}}\sqrt{\sqrt{M_1(v,a;\alpha,\beta)^2+b(v,a;\alpha,\beta)^2}+M_1(v,a;\alpha,\beta)}\r\} \\
&     \hspace{5mm}\times\exp\l\{-(T-t)\frac{(-h)\delta\alpha}{\sqrt{2}}\sqrt{\sqrt{M_1(v,a;\alpha,1+\beta)^2+b(v,a;\alpha,1+\beta)^2}+M_1(v,a;\alpha,1+\beta)}\r\} \\
&\leq C(t)\exp\l\{-(T-t){(1+h)\delta\alpha}\sqrt{M_1(v,a;\alpha,\beta)}\r\}\exp\l\{-(T-t){(-h)\delta\alpha}\sqrt{M_1(v,a;\alpha,1+\beta)}\r\} \\
&=    C(t)\exp\l\{-(T-t)\delta\l((1+h)\sqrt{v^2+\alpha^2-(a+\beta)^2}+(-h)\sqrt{v^2+\alpha^2-(a+1+\beta)^2}\r)\r\} \\
&\leq C(t)\exp\{-(T-t)\delta v\}.
\end{align*}
Note that the last inequality follows from the fact that $\alpha^2-(a+\beta)^2>0$ and $\alpha^2-(a+1+\beta)^2>0$ hold by Assumption \ref{ass0}.
\fin

\begin{lem}\label{lem-A3}
For any $v\in[0,\infty)$ and any $a\in(\frac{3}{2},2]$,
\[
|W(v,a+1)-W(v,a)| \leq \sqrt{2}\delta\l(v+\sqrt{\alpha^2-(a+\beta)^2+2(a+1+\beta)^2}\r)
\]
holds.
\end{lem}

\proof
Denoting $M_1':=M_1(v,a+1)$, $b':=b(v,a+1)$, $M_1:=M_1(v,a)$ and $b:=b(v,a)$ for short, we have
\begin{align}
&|W(v,a+1)-W(v,a)| \notag \\
&=    \frac{\delta\alpha}{\sqrt{2}}\Bigg|i\l(\sqrt{\sqrt{M_1'^2+b'^2}-M_1'}-\sqrt{\sqrt{M_1^2+b^2}-M_1}\r)-\sqrt{\sqrt{M_1'^2+b'^2}+M_1'}+\sqrt{\sqrt{M_1^2+b^2}+M_1}\Bigg| \notag \\
&\leq \delta\alpha\sqrt{\sqrt{M_1'^2+b'^2}+\sqrt{M_1^2+b^2}}.
\label{eq4:3:6}
\end{align}
Since $a+\beta>0$, we have
\[
M_1-M_1' = \frac{1}{\alpha^2}\l((a+1+\beta)^2-(a+\beta)^2\r)>0
\]
and
\[
b'^2-b^2=\frac{4v^2}{\alpha^4}\l((a+1+\beta)^2-(a+\beta)^2\r)>0,
\]
which imply that
\begin{align}
\mbox{(\ref{eq4:3:6})}
&\leq \delta\alpha\sqrt{2{\sqrt{M_1^2+b'^2}}} = \sqrt{2}\delta\sqrt[4]{(v^2+\alpha^2-(a+\beta)^2)^2+4v^2(a+1+\beta)^2} \notag \\
&=    \sqrt{2}\delta\sqrt[4]{v^4+2v^2(\alpha^2-(a+\beta)^2+2(a+\beta+1)^2)+(\alpha^2-(a+\beta)^2)^2}.
\label{eq4:3:8}
\end{align}
Setting
\[
\l\{
\begin{array}{l}
p := \alpha^2-(a+\beta)^2+2(a+\beta+1)^2, \\
q := p^2-(\alpha^2-(a+\beta)^2)^2, \\
\end{array}\r.
\]
we have $p>0$ and $q>0$ for any $a\in(\frac{3}{2},2]$ by Assumption \ref{ass0}; and
\[
\mbox{(\ref{eq4:3:8})} = \sqrt{2}\delta\sqrt[4]{(v^2+p)^2-q} \leq \sqrt{2}\delta\sqrt{v^2+p} \leq \sqrt{2}\delta(v+\sqrt{p}).
\]
This completes the proof of Lemma \ref{lem-A3}.
\fin

\noindent
{\it Proof of Proposition \ref{prop3}.} \ 
Firstly, Lemma \ref{lem-A3} implies that
\begin{align}
&\l|\frac{1}{\pi}\int_w^\infty K^{-iv-a+1}\Big(W(v,a+1)-W(v,a)-W(0,1)\Big)\frac{\phi_{T-t}(iv-a)S_{t-}^{iv+a}}{(iv+a)(iv+a-1)}dv\r| \notag \\
&\leq \frac{1}{\pi}\int_w^\infty\l|K^{-iv-a+1}\r|\Big(|W(v,a+1)-W(v,a)|+|W(0,1)|\Big)\l|\frac{\phi_{T-t}(iv-a)S_{t-}^{iv+a}}{(iv+a)(iv+a-1)}\r|dv \notag \\
&\leq \frac{{\delta}K^{-a+1}}{\pi}\int_w^\infty\l(\sqrt{2}(v+\sqrt{p})+\sqrt{2}\r)\l|\frac{\phi_{T-t}(iv-a)S_{t-}^{iv+a}}{(iv+a)(iv+a-1)}\r|dv,
\label{eq4:3:14}
\end{align}
where $p$ is defined in the proof of Lemma \ref{lem-A3}.
Remark that the last inequality in (\ref{eq4:3:14}) holds since
\begin{align*}
|W(0,1)| &=    \delta\alpha\l(\sqrt{M_1(0,1)}-\sqrt{M_2}\r) = \delta\alpha\l(\sqrt{M_2-\frac{1+2\beta}{\alpha^2}}-\sqrt{M_2}\r) \\
         &\leq \delta\alpha\sqrt{-\frac{1+2\beta}{\alpha^2}} \leq \sqrt{2}\delta
\end{align*}
by Assumption \ref{ass0}.
Now, note that
\begin{align*}
|(iv+a-1)(iv+a)| &= \sqrt{(a^2-a-v^2)^2+(2a-1)^2v^2} \\
                 &= \sqrt{v^4+(2a^2-2a+1)v^2+(a^2-a)^2} \geq v^2.
\end{align*}
Thus, Proposition \ref{prop-A1} implies that
\begin{align*}
\l|\frac{\phi_{T-t}(iv-a)S_{t-}^{iv+a}}{(iv+a)(iv+a-1)}dv\r| \leq \frac{S^a_{t-}C(t)}{v^2}e^{-(T-t)\delta v}.
\end{align*}
As a result, noting that $w>1$, we obtain
\begin{align*}
\mbox{(\ref{eq4:3:14})}
&\leq \frac{\delta K^{-a+1}S^a_{t-}C(t)}{\pi}\int_w^\infty\l(\sqrt{2}(v+\sqrt{p})+\sqrt{2}\r)\frac{1}{v^2}e^{-(T-t)\delta v}dv \\
&\leq \frac{\delta K^{-a+1}S^a_{t-}C(t)}{\pi}\int_w^\infty\l(2\sqrt{2}+\sqrt{2p}\r)e^{-(T-t)\delta v}dv \\
&=    \frac{K^{-a+1}S^a_{t-}C(t)}{\pi}\frac{\sqrt{2}(2+\sqrt{p})}{T-t}e^{-(T-t)\delta w}.
\end{align*}
This completes the proof of Proposition \ref{prop3}.
\fin



\begin{thebibliography}{00}
\bibitem{AI} Arai, T., Imai, Y.:
   A closed-form representation of mean-variance hedging for additive processes via Malliavin calculus, preprint. Available at \url{https://arxiv.org/abs/1702.07556}
\bibitem{AIS}
   Arai, T., Imai, Y., Suzuki, R.:
   Numerical analysis on local risk-minimization for exponential L\'evy models, International Journal of Theoretical and Applied Finance, Vol.19, 1650008, (2016)
\bibitem{AS}
   Arai, T., Suzuki, R.:
   Local risk-minimization for L\'evy markets, International Journal of Financial Engineering, Vol.2, 1550015, (2015)
\bibitem{BN}
   Barndorff-Nielsen, O.E.:
   Normal inverse Gaussian processes and the modelling of stock returns. Aarhus Universitet. Department of Theoretical Statistics. (1995)
\bibitem{BN0}
   Barndorff-Nielsen, O.E.:
   Processes of normal inverse Gaussian type. Finance and Stochastics, Vol.2, 41-68  (1997)
\bibitem{BN1}
   Barndorff-Nielsen, O.E.:
   Normal inverse Gaussian distributions and stochastic volatility modelling. Scandinavian Journal of statistics, Vol.24, 1-13 (1997)
\bibitem{BB}
    Benth, F.E., \v{S}altyt\.e-Benth, J.:
    The normal inverse Gaussian distribution and spot price modelling in energy markets, International Journal of Theoretical and Applied Finance, Vol.7, 177-192 (2004)
\bibitem{CM}
   Carr, P., Madan, D.:
   Option valuation using the fast Fourier transform., Journal of computational Finance, Vol.2, 61-73 (1999).
\bibitem{CT}
   Cont, R., Tankov, P.:
   Financial Modelling with Jump Process, Chapman \& Hall, London. (2004)
\bibitem{DI}
   Delong, \L., Imkeller, P.:
   On Malliavin's differentiability of BSDEs with time delayed generators driven by Brownian motions and Poisson random measures.
   Stochastic Process. Appl., Vol.120, 1748-1775 (2010)
\bibitem{R1}
   Rydberg, T.H.:
   The normal inverse Gaussian L\'evy process: simulation and approximation. Communications in statistics. Stochastic models, Vol.13, 887-910 (1997)
\bibitem{R2}
   Rydberg, T.H.:
   A note on the existence of unique equivalent martingale measures in a Markovian setting. Finance and Stochastics, Vol.1, 251-257 (1997)
\bibitem{Scho}
   Schoutens, W.:
   L\'evy Process in Finance: Pricing Financial Derivatives, Hoboken: John Wiley \& Sons. (2003)
\bibitem{Sch}
   Schweizer, M.:
   A guided tour through quadratic hedging approaches. Option pricing, interest rates and risk management., Jouini E., Cvitanic J. and Musiela M. editors.
   Cambridge University Press, 538--574, (2001)
\bibitem{Sch3}
   Schweizer, M.:
   Local risk-minimization for multidimensional assets and payment streams. Banach Cent. Publ, Vol.83, 213-229 (2008)
\bibitem{S07}
    Sol\'e, J.L., Utzet F., Vives, J.:
    Canonical L\'evy process and Malliavin calculus, Stochastic Process. Appl., Vol.117, 165-187 (2007)
\end{thebibliography}
\end{document}